\documentclass[journal]{IEEEtran}
\usepackage[latin9]{inputenc}
\usepackage{array}
\usepackage{url}
\usepackage{amsmath}
\usepackage{amssymb}
\usepackage{graphicx}
\PassOptionsToPackage{normalem}{ulem}
\usepackage{ulem}

\makeatletter

\providecommand{\tabularnewline}{\\}

\usepackage{amsfonts}
\usepackage{multirow}
\usepackage{url}
\providecommand{\U}[1]{\protect \rule{.1in}{.1in}}

\makeatother

\begin{document}
\title{Simulation Credibility Assessment Methodology with FPGA-based Hardware-in-the-loop
Platform}

\author{Xunhua Dai, Chenxu Ke, Quan Quan and Kai-Yuan Cai \thanks{The authors are with School of Automation Science and Electrical Engineering,
		Beihang University, Beijing 100191, China.}}
\maketitle
\begin{abstract}
Electronic control systems are becoming more and more complicated,
which makes it difficult to test them sufficiently only through experiments.
Simulation is an efficient way in the development and testing of complex
electronic systems, but the simulation results are always doubtful
by people due to the lack of credible simulation platforms and assessment
methods. This paper proposes a credible simulation platform based
on real-time FPGA-based hardware-in-the-loop (HIL) simulation, and
then an assessment method is proposed to quantitatively assess its
simulation credibility. By using the FPGA to simulate all sensor chips,
the simulation platform can ensure that the tested electronic system
maintains the same hardware and software operating environment in
both simulations and experiments, which makes it possible to perform
the same tests in the simulation platform and the real experiment
to compare and analyze the simulation errors. Then, the testing methods
and assessment indices are proposed to assess the simulation platform
from various perspectives, such as performance, time-domain response,
and frequency-domain response. These indices are all normalized to
the same scale (from 0 to 1) and mapped to a uniform assessment criterion,
which makes it convenient to compare and synthesize different assessment
indices. Finally, an overall assessment index is proposed by combining
all assessment indices obtained from different tests to assess the
simulation credibility of the whole simulation platform. The simulation
platform and the proposed assessment method are applied to a multicopter
system, where the effectiveness and practicability are verified by
simulations and experiments.
\end{abstract}


\section{Introduction}

Currently, electronic control systems (e.g., autopilot systems of
unmanned vehicles) are becoming more and more complicated, which makes
it more and more difficult to test them sufficiently only through
experiments. Therefore, simulation techniques, especially hardware-in-the-loop
(HIL) simulations, are more and more widely used in the development
and testing phases of complex electronic control systems, such as
power systems \cite{Mao2018,Hadizadeh2019}, aircraft systems \cite{Qi2017},
automotive systems \cite{chen2018autonomous}, and robotic systems
\cite{Tejado2016}. Although experiments are considered to be more
trusted than simulation tests, for many high-complex electronic control
systems (e.g., autopilot systems of unmanned aircraft), comprehensive
experimental testing is usually high-cost, inefficient, dangerous
and regulatory restricted \cite{Shi2017}. With the ever-increasing
safety requirements of electronic control systems, the experimental
testing methods become increasingly inefficient in revealing potential
safety issues and covering critical test cases. Besides, in experiments,
the true states of a plant can only be estimated by external measuring
devices or onboard sensors whose measured results may be easily affected
by the many uncontrollable factors, such as noise, vibration, temperature,
and unexpected interference or failure. Instead, in simulations, the
true states are known precisely, which makes it more efficient and
accurate in assessing the performance and safety level of an electronic
control system. However, the simulation credibility \cite{mehta2016simulation}
is still the most concerned problem for people (e.g., users, companies,
and certification authorities) to acknowledge that the simulation
results can be as credible as real experiments.

According to \cite{mehta2016simulation}, simulation credibility can
be assessed both qualitatively and quantitatively. The former assesses
the quality of simulation by professional engineers based on circumstantial
evidence, which is simple but not convincing enough; the latter requires
to quantify the simulation errors relative to real systems, which
is convincing but usually difficult to implement. In practice, the
qualitative assessment is widely adopted in simulation credibility
analysis. For example, in \cite{Mai2017,Roinila2019}, the credibility
of the HIL simulation platforms is assessed by qualitatively comparing
the simulations results with experimental results from several aspects.
Since there is no widely accepted index and standard for simulation
credibility assessment, it is hard to quantitatively compare and assess
different simulation platforms from an objective and comprehensive
perspective. For the above concerns, a comprehensive survey for the
verification and validation of simulation credibility is introduced
in \cite{Sargent2017}, where the following problems are revealed.
(i) Traditional simulations are too separated from the actual hardware
system, which makes it difficult to compare the simulation results
with the experimental results. (ii) The simulation credibility is
informal and not accurate enough because it is mainly assessed by
the experience of engineers \cite{Morrison2019}. In summary, the
simulation credibility should be ensured from two aspects. (i) The
credibility of the simulation platform should be first guaranteed
by maintaining the same hardware and software operating environment
of the tested electronic control systems in both simulations and experiments.
(ii) A qualitative credibility assessment method should be proposed
to assess the simulation results from a more objective and comprehensive
perspective.

Electronic control systems can usually be divided into the plant system
and the control system. In software simulation, the plant simulation
software runs on the same computer with the control algorithms, which
is different from the real system whose algorithms usually run in
specialized hardware. As a result, the software simulation results
are usually considered to be less credible compared with experiments
in real systems. Then, HIL simulation is proposed to increase the
simulation credibility by using Real-Time (RT) simulation computers
and real control system in simulations. However, limited by the performance
of RT simulation computers, it is usually difficult for traditional
RT simulation computers to simulate some sensors with high-speed communication
interfaces or high-frequency analog circuits \cite{lucia2010real}.
For example, a nanosecond-level RT update frequency is required to
simulate the high-speed Serial Peripheral Interface (SPI) communication,
which is a difficult task for traditional RT computers (model update
frequency usually smaller 100kHz) with commercial Central Processing
Units (CPUs) \cite{saad2015real}. In recent years, the Field Programmable
Gate Array (FPGA) \cite{matar2010fpga} is becoming a standard component
for RT simulation computers, and Commercial-Off-The-Shelf (COTS) RT
simulation computers (such as RT-LAB and NI-PXI) start to have the
ability to directly simulate electronic chips and circuits with a
nanosecond-level RT update frequency \cite{Mikkili2015,Noureen2018}.
Based on this, FPGA-based HIL simulation platforms can simulate almost
everything (including plant motion, environment conditions, sensor
hardware, and interfaces) outside the control system. By using the
same control system in both HIL simulations and experiments, the structure
difference between simulation systems and real systems can be significantly
controlled.

In \cite{mehta2016simulation}, several assessment methods are proposed
to assess simulation credibility, but these methods mainly focus on
one specific feature instead of the whole system. Besides, many studies
\cite{Mai2017,Roinila2019} use the simulation errors (result error
between simulation and experiment) as assessment indices to assess
the simulation accuracy, but these indices are usually of a range
from 0 to $+\infty$, which are not as convenient as normalized indices
with a range from 0 to 1. Besides, different assessment indices may
have different physics meaning, scales, and units, so it is difficult
to combine different indices for comprehensively assessing the whole
simulation system. For example, \cite{remple2006aircraft} proposes
a cost function $J\in(0,+\infty]$ to assess the modeling accuracy
by analyzing the Bode magnitude and phase information in the frequency
domain. The cost function $J$ is obtained by combining the magnitude
error and the phase error (between simulations and experiments) with
a constant scaling factor determined by human experience. One disadvantage
of using constant factors to combining different indices with value
range $(0,+\infty]$ is that some indices will be ignored when their
orders of magnitudes are too different, which requires people to find
appropriate scaling factors for specific systems. In summary, there
is still a lack of practical and comprehensive simulation assessment
methods widely recognized and accepted in the simulation filed.

The main work and contributions of this paper are as follows. (i)
An FPGA-based HIL simulation platform is proposed to be able to simulate
all situations as real experiments do and eliminate disturbance factors
for simulation credibility assessments. (ii) Normalized assessment
indices (the index range is from 0 to 1) are proposed and mapped into
a uniform assessment criterion (e.g., a passing mark 0.6), which are
practical and intuitive for comparison between different physical
quantities. (iii) Multiple factors (including performance, time-domain
response, and the frequency-domain response) are considered to assess
the simulation credibility of the HIL platform comprehensively. (iv)
An overall assessment index is proposed by combining the above indices
to assess the simulation credibility of the whole HIL simulation system.
In the verification part, the HIL simulation platform is successfully
applied to a quadcopter system, and its simulation results are compared
with the experimental results to assess the simulation credibility
with the proposed method. The experiments and comparisons demonstrate
the effectiveness and practicability of the proposed platform and
the credibility assessment method.

The rest of the paper is organized as follows. \textsl{Section }\ref{sec:2}
gives a description of the FPGA-based HIL simulation platform and
the testing methods for obtaining the simulation errors. Then, the
simulation credibility assessment method is presented in \textsl{Section
}\ref{sec:3}. In \textsl{Section }\ref{sec:4}, the proposed platform
and the assessment method are applied to a multicopter system to verify
the proposed methods with experiments. \textsl{Section }\ref{sec:5}
presents the conclusion and future work.

\section{HIL Platform and Testing Method}

\label{sec:2}

\subsection{FPGA-based HIL Simulation}

\begin{figure}[tbh]
\centering \includegraphics[width=0.45\textwidth]{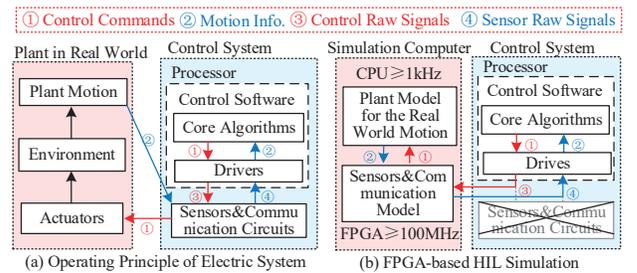}\caption{Testing methods in real-world experiments and the FPGA-based HIL simulation
platform.}
\label{Fig02}
\end{figure}

A modern complex electronic control system (e.g., autonomous vehicles
and aircraft) can be divided into the plant system (e.g., the vehicle
body and the actuators) and the control system (e.g., the autopilot
system), where the control system is the most important component
that determines the performance in normal situation and safety in
failure situations. Fig.\,\ref{Fig02}(a) presents the operating
principle of a real electronic control system, where the plant motion
information is sensed by sensors and then transmitted into the control
system to acquire control commands for driving the actuators. In order
to maximally maintain the system structure and operating environment
as the real electronic control system, an FPGA-based HIL simulation
method is proposed with the structure depicted in Fig.\,\ref{Fig02}(b),
where the sensors and communication interfaces are blocked and replaced
by a model running in the FPGA to exchange the simulated sensor data
and control signals with the control system. With comprehensively
modeling all necessary elements (e.g., the plant system, sensors,
environment, measuring noise, disturbances, and faults), the HIL simulation
platform can theoretically simulate any situation as the electronic
control system.

\subsection{Testing and Assessment Framework}

\begin{figure}
\centering \includegraphics[width=0.45\textwidth]{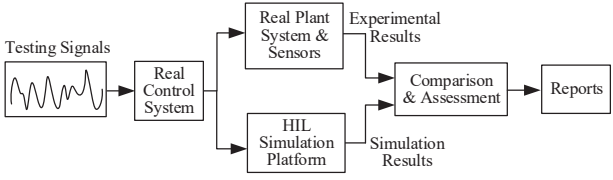}\caption{Simulation validation testing structure.}
\label{Fig08}
\end{figure}

As shown in Fig.\,\ref{Fig08}, the most effective way to assess
the simulation credibility is to input the same signals to both the
HIL simulation system and the real system to compare their result
errors. The input signals should be selected from multiple aspects
(including system performance, time-domain response, and frequency-domain
response) to fully excite the system to reveal the system properties
comprehensively. By using the same control system (see Fig.\,\ref{Fig08})
in both simulations and experiments, the simulation errors caused
by hardware and software differences of control systems can be controlled
to the utmost extent, which significantly improves the credibility
of the simulation platform compared with other simulation methods.

\subsection{Assessment Index Normalization}

The error $e$ between simulation and experimental is the most important
index to assess the simulation credibility. However, its value range
$e\in[0,+\infty)$ is not suitable for comparison and assessment.
In practice, an error threshold $\varepsilon>0$ can be obtained from
design experience or related standards to define the accepting error
range $e\leq\varepsilon$ for assessment. Based on that, a normalization
function is introduced here to map the error index $e\in[0,+\infty)$
to an assessment index $\eta\in(0,1]$ with the error bound $e\leq\varepsilon$
corresponding to a desired passing mark $\eta\geq\eta_{\text{pass}}$
as

\begin{equation}
\eta\triangleq f_{\text{norm}}\left(e,\varepsilon\right)=\frac{K_{\text{e}}\cdot\varepsilon}{\sqrt{\left(K_{\text{e}}\cdot\varepsilon\right)^{2}+e^{2}}}\label{eq:NormIndex}
\end{equation}
where the coefficient $K_{\text{e}}\in\mathbb{R^{+}}$ is a scale
factor to ensure $\eta_{\text{pass}}=f_{\text{norm}}\left(e=\varepsilon,\varepsilon\right)$,
which gives
\begin{equation}
K_{\text{e}}=\frac{\eta_{\text{pass}}}{\sqrt{1-\eta_{\text{pass}}^{2}}}.\label{eq:passingline}
\end{equation}
Noteworthy, to accord with people's assessing habits, the passing
mark $\eta_{\text{pass}}$ can usually be selected as $\eta_{\text{pass}}\triangleq0.6$
(or marked with 60\%), which yields from (\ref{eq:passingline}) that
$K_{\text{e}}=0.75$. The physical meaning for the assessment index
$\eta$ is that: $\eta\rightarrow0$ for worse simulation credibility,
$\eta\rightarrow1$ for better credibility, and $\eta=0.6$ for a
credibility passing line.

\section{Simulation Credibility Assessment}

\label{sec:3}

\begin{figure}
\centering \includegraphics[width=0.45\textwidth]{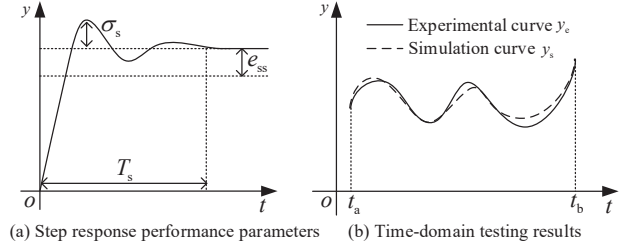}\caption{Typical simulation and experimental results.}
\label{Fig08-1}
\end{figure}

The characteristics of an electronic control system can usually be
described by many performance parameters, such as endurance, response
delay, and maximum speed. Meanwhile, many testing results can also
be summarized by several performance parameters, such as the percent
overshoot $\sigma_{\text{s}}$, the settling time $T_{\text{s}}$
of a step response curve presented in Fig.\,\ref{Fig08-1}(a). In
practice, comparing the performance parameters of the simulation system
with the real system is the most commonly used way to assess simulation
credibility, but it may ignore many important dynamic or frequency
features. Thus, the time-domain testing and frequency-domain testing
should also be considered in the simulation assessment method. This
section will assess the simulation credibility of the whole simulation
system by considering the above features comprehensively.

\subsection{Performance Credibility}

By applying direct measurement methods \cite{quan2017introduction}
or system identification methods \cite{remple2006aircraft} to the
HIL simulation system and the real system presented in Fig.\,\ref{Fig08},
the performance parameters can be obtained for the simulation system
$p_{\text{s}}$ and the real experimental system $p_{\text{e}}$,
and the simulation error $e_{\text{p}}$ is defined as
\begin{equation}
e_{\text{p}}\triangleq\left|p_{\text{e}}-p_{\text{s}}\right|.\label{eq:Ekey}
\end{equation}
As mentioned in (\ref{eq:NormIndex}), an error threshold $\varepsilon_{\text{p}}\in\mathbb{R^{+}}$
should be obtained from design experience or related standards for
the assessment requirements. To simplify the selection process for
$\varepsilon_{\text{p}}$ and minimize the human subjectively, a dynamic
selection method for $\varepsilon_{\text{p}}$ is proposed in this
paper as
\begin{equation}
\varepsilon_{\text{p}}=K_{\text{p}}\cdot\left|p_{\text{e}}\right|.\label{eq:thres}
\end{equation}
where $K_{\text{p}}\in\mathbb{R^{+}}$ is a percentage coefficient,
and $K_{\text{p}}=5\%$ is usually applicable for most situations.
The expression of (\ref{eq:thres}) indicates the threshold $\varepsilon_{\text{p}}$
is dynamically adjusted with the detailed experimental value $p_{\text{e}}$.
This is reasonable because a larger measured value usually has a larger
error bound. The percentage coefficient $K_{\text{p}}$ is also adjustable
according to the actual situation. For example, a larger coefficient
$K_{\text{p}}$ should be selected when the disturbance or measuring
errors are relatively large. Noteworthy, (\ref{eq:thres}) may not
apply to the situation $p_{\text{e}}=0$ because $\varepsilon_{\text{p}}>0$
must be satisfied for the following computation. In this case, other
methods should be applied to determine $\varepsilon_{\text{p}}$,
such as $\varepsilon_{\text{p}}=K_{\text{p}}\cdot\left|p_{\text{s}}\right|$.

Since $e_{\text{p}}\in[0,+\infty)$ is suitable for credibility assessment,
the normalization function in (\ref{eq:NormIndex}) is applied to
define the performance credibility index $\eta_{\text{p}}$ as
\begin{equation}
\eta_{\text{p}}=f_{\text{norm}}\left(e_{\text{p}},\varepsilon_{\text{p}}\right)=\frac{K_{\text{e}}\cdot\varepsilon_{\text{p}}}{\sqrt{\left(K_{\text{e}}\cdot\varepsilon_{\text{p}}\right)^{2}+e_{\text{p}}^{2}}}\label{eq:AAA}
\end{equation}
where, similar to (\ref{eq:NormIndex}), the range of $\eta_{\text{p}}$
is $(0,1]$ with a the passing mark $\eta_{\text{p}}\geq0.6$ (corresponds
to $e_{\text{p}}\leq\varepsilon_{\text{p}}$) to present the matching
degree with the real system.

\subsection{Time-domain Credibility}

The time-domain testing results obtained from systems in Fig.\,\ref{Fig08}
can be described by the data curves presented in Fig.\,\ref{Fig08-1}(b),
where $y_{\text{e}}\left(t\right)$ denotes the experimental curve,
$y_{\text{s}}\left(t\right)$ denotes the simulation curve, and $t\in\left[t_{\text{a}},t_{\text{b}}\right]$
denotes the tested interval. The time-domain credibility can be assessed
by obtaining the average error between the simulation curves and the
experimental curves. First, dividing the interval $\left[t_{\text{a}},t_{\text{b}}\right]$
into $n_{\text{t}}$ sample points as $t_{1},t_{2},\cdots,t_{n_{\text{t}}}$,
the average error between the simulation curve and the experimental
curve can be computed by
\begin{equation}
e_{\text{t}}=\sqrt{\frac{1}{n_{\text{t}}}\sum_{1}^{n_{\text{t}}}\left(y_{\text{e}}\left(t_{i}\right)-y_{\text{s}}\left(t_{i}\right)\right)^{2}}.\label{eq:RQ}
\end{equation}
 Secondly, similar to (\ref{eq:thres}), a feasible selection method
for the error threshold $\varepsilon_{\text{t}}$ is proposed as
\begin{equation}
\varepsilon_{\text{t}}=K_{\text{p}}\cdot\max_{1\leq i,j\leq n_{\text{t}}}\left|y_{\text{e}}\left(t_{i}\right)-y_{\text{e}}\left(t_{j}\right)\right|\label{eq:EtThre}
\end{equation}
where $K_{\text{p}}$ is a percentage coefficient as defined in (\ref{eq:thres}).
The expression of (\ref{eq:EtThre}) indicates the maximum tolerable
threshold $\varepsilon_{\text{t}}$ is proportional to the maximum
range of the experimental curve $y_{\text{e}}\left(t\right)$. Finally,
according to (\ref{eq:NormIndex}), the normalized time-domain credibility
index $\eta_{\text{t}}$ is given by
\begin{equation}
\eta_{\text{t}}=f_{\text{norm}}\left(e_{\text{t}},\varepsilon_{\text{t}}\right).\label{eq:CTime}
\end{equation}
Noteworthy, the variable $t$ for the curve functions $y_{\text{e}}\left(t\right)$
and $y_{\text{s}}\left(t\right)$ does not have to be time. Any measured
curves (e.g., movement trajectories, motor throttle-speed curves,
and air resistance curves) can be applied to assess the time-domain
credibility $\eta_{\text{t}}$ of the simulation systems. To reduce
the calculation error, curve smoothing methods should be applied to
(\ref{eq:RQ}) when the obtained curves are affected by disturbances
or measuring noises. Besides, the time-domain index $\eta_{\text{t}}$
is not suitable for assessing stochastic curves (e.g., noise signals
and vibration signals), which can be assessed by the performance assessment
index $\eta_{\text{p}}$ with statistical parameters (e.g., mean value
and variance).

\subsection{Frequency-domain Credibility}

\subsubsection{Sweep-frequency Result Coherence}

The frequency-domain testing should also be performed for systems
in Fig.\,\ref{Fig08} to sufficiently excite the system characteristics
within the frequency range of interest. First, according to \cite{remple2006aircraft},
a coherence index $\eta_{\text{\ensuremath{\eta}o}}$ at the given
frequency point $f$ is necessary for evaluating the frequency-domain
test results as

\begin{equation}
\eta_{\text{co}}\left(f\right)=\frac{\left|\hat{G}_{xy}\left(f\right)\right|^{2}}{\left|\hat{G}_{xx}\left(f\right)\right|\cdot\left|\hat{G}_{yy}\left(f\right)\right|}\label{eq:bbb}
\end{equation}
where $\hat{G}_{xy}\left(f\right)$ is the cross-spectrum estimation
of the input signal and the output signal at the frequency point $f$,
$\hat{G}_{xx}\left(f\right)$ is the auto-spectrum estimation of the
input signal, and $\hat{G}_{yy}\left(f\right)$ is the auto-spectrum
estimation of the output signal \cite[p. 30]{remple2006aircraft}.
The range of the coherence index $\eta_{\text{co}}$ is $(0,1]$,
and $\eta_{\text{co}}\rightarrow1$ denotes the results obtained by
frequency-domain testing are more accurate and credible. For the given
frequency range of interest $f_{\text{min}}\leq f\leq f_{\text{max}}$,
only when the following criterion is satisfied
\begin{equation}
\eta_{\text{co}}\left(f\right)\geq\varepsilon_{\text{co}},\text{ }f_{\text{a}}\leq f\leq f_{\text{b}}\label{eq:Cohe}
\end{equation}
then the frequency-domain testing results are considered accurate
and credible for the following assessment process, where a threshold
$\varepsilon_{\text{co}}=0.6$ is recommended in \cite[p. 38]{remple2006aircraft}.

\subsubsection{Simulation Errors in Magnitude and Phase Plots}

\begin{figure}
\centering \includegraphics[width=0.45\textwidth]{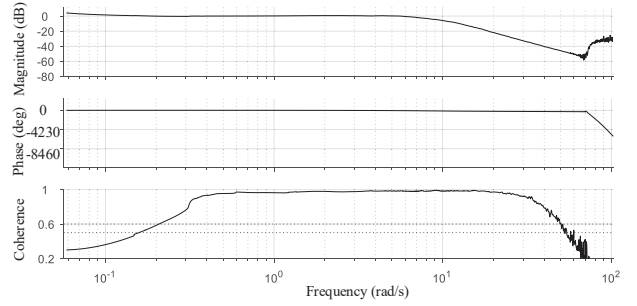}\caption{Sweep frequency data processing results with the software CIFER$^{\circledR}$.}
\label{Fig08-1-1}
\end{figure}

The frequency-domain response of a system can be described by Bode
plots, which include the magnitude plot and the phase plot. By performing
sweep-frequency tests to the real system and the simulation system
in Fig.\,\ref{Fig08}, the magnitude and phase curves can be obtained
by frequency-domain identification tools such as CIFER$^{\circledR}$
\cite{remple2006aircraft} and MATLAB$^{\circledR}$. In this paper,
the CIFER$^{\circledR}$ software is applied to process the sweep-frequency
testing results, and a demo of obtained results is depicted in Fig.\,\ref{Fig08-1-1}.

Let $M_{\text{e}}\left(f\right)$ and $P_{\text{e}}\left(f\right)$
be the experimental magnitude and phase curves from the real system,
and $M_{\text{s}}\left(f\right)$ and $P_{\text{s}}\left(f\right)$
be the simulated magnitude and phase curves from the simulation system.
Then, by dividing the frequency interval $\left[f_{\text{a}},f_{\text{b}}\right]$
to $n_{f}$ sample points $f_{1},f_{2},\cdots f_{n_{f}}$, the average
magnitude curve error $e_{\text{mag}}$ and phase curve $e_{\text{pha}}$
can be obtained as
\begin{align}
e_{\text{mag}} & =\sqrt{\frac{1}{n_{\text{f}}}\sum_{1}^{n_{\text{f}}}W_{\gamma}^{2}\left(f_{i}\right)\left(M_{\text{e}}\left(f_{i}\right)-M_{\text{s}}\left(f_{i}\right)\right)^{2}}\label{eq:emag}\\
e_{\text{pha}} & =\sqrt{\frac{1}{n_{\text{f}}}\sum_{1}^{n_{\text{f}}}W_{\gamma}^{2}\left(f_{i}\right)\left(P_{\text{e}}\left(f_{i}\right)-P_{\text{s}}\left(f_{i}\right)\right)^{2}}\label{eq:Epha}
\end{align}
where $W_{\gamma}\left(f\right)\in(0,1]$ is a weighting function
($\eta_{\text{co}}\rightarrow1\Rightarrow W_{\gamma}\rightarrow1$)
to ensure the sample points $f_{i}$ with higher coherence $\eta_{\text{co}}$
have larger weight $W_{\gamma}$. Based on the research in \cite[p. 280]{remple2006aircraft},
the weighting function $W_{\gamma}\left(f\right)$ is given by
\begin{equation}
W_{\gamma}\left(f\right)=\frac{\left(1-\text{e}^{-\eta_{\text{co}}\left(f\right)}\right)}{1-\text{e}^{-1}}
\end{equation}
which ensures the most effectively use of the testing data with different
testing reliability. Noteworthy, $W_{\gamma}\left(f\right)\equiv1$
can be applied to simplify the computational process of (\ref{eq:emag})
and (\ref{eq:Epha}) when accuracy requirement is not too high.

\subsubsection{Frequency-domain Assessment Index}

Let $\varepsilon_{\text{mag}}\in\mathbb{R^{+}}$ and $\varepsilon_{\text{pha}}\in\mathbb{R^{+}}$
present the thresholds for the magnitude and phase average errors
$e_{\text{mag}}$ and $e_{\text{pha}}$, receptively. Similar to (\ref{eq:EtThre}),
the selection methods for $\varepsilon_{\text{mag}}$ and $\varepsilon_{\text{pha}}$
are given by
\begin{align}
\varepsilon_{\text{mag}} & =K_{\text{p}}\cdot\max_{1\leq i,j\leqslant n_{\text{f}}}\left|M_{\text{e}}\left(f_{i}\right)-M_{\text{e}}\left(f_{j}\right)\right|\label{eq:EmagPha}\\
\varepsilon_{\text{pha}} & =K_{\text{p}}\cdot\max_{1\leq i,j\leqslant n_{\text{f}}}\left|P_{\text{e}}\left(f_{i}\right)-P_{\text{e}}\left(f_{j}\right)\right|\label{eq:EmagPha1}
\end{align}
where $K_{\text{p}}$ is a percentage coefficient as defined in (\ref{eq:thres}).

Letting $\eta_{\text{mag}}\in(0,1]$ and $\eta_{\text{pha}}\in(0,1]$
present the model credibility in the magnitude aspect and phase aspect,
their expressions can be obtained by (\ref{eq:NormIndex}) as
\begin{equation}
\begin{array}{c}
\eta_{\text{mag}}=f_{\text{norm}}\left(e_{\text{mag}},\varepsilon_{\text{mag}}\right)\\
\eta_{\text{pha}}=f_{\text{norm}}\left(e_{\text{pha}},\varepsilon_{\text{pha}}\right)
\end{array}\label{eq:MatFre}
\end{equation}
Finally, the overall frequency-domain credibility index $\eta_{\text{f}}\in(0,1]$
is combined from (\ref{eq:MatFre}) as
\begin{equation}
\eta_{\text{f}}=\sqrt{\frac{1}{2}\left(\eta_{\text{mag}}^{2}+\eta_{\text{pha}}^{2}\right)}\label{eq:freqIndex}
\end{equation}
where $\eta_{\text{f}}$ is capable of combining the errors $e_{\text{mag}}$
and $e_{\text{pha}}$ at the same scale, and $\eta_{\text{f}}$ is
also normalized index with a passing mark 0.6 as (\ref{eq:NormIndex}).

\subsection{Overall Simulation Credibility}

Assuming that enough assessment tests ($n_{\text{p}}$ performance
parameter tests $\eta_{\text{p},i}$, $n_{\text{t}}$ time-domain
tests $\eta_{\text{t},i}$, and $n_{\text{\text{f}}}$ frequency-domain
tests $\eta_{\text{\text{f}},i}$) have been performed with the whole
assessment indices for the performance credibility $\overline{\eta}_{\text{p}}$,
the time-domain credibility $\overline{\eta}_{\text{t}}$, and the
frequency-domain credibility $\overline{\eta}_{\text{f}}$ are given
by
\begin{equation}
\overline{\eta}_{\text{p}}=\sqrt{\frac{1}{n_{\text{p}}}\sum_{i=1}^{n_{\text{p}}}\eta_{\text{p},i}^{2}},\,\overline{\eta}_{\text{t}}=\sqrt{\frac{1}{n_{\text{t}}}\sum_{i=1}^{n_{\text{t}}}\eta_{\text{t},i}^{2}},\,\overline{\eta}_{\text{f}}=\sqrt{\frac{1}{n_{\text{f}}}\sum_{i=1}^{n_{\text{\text{f}}}}\eta_{\text{\text{f}},i}^{2}}.\label{eq:aveP}
\end{equation}
Then, the overall assessment index $\eta_{\text{all}}$ for the whole
system is given by
\begin{equation}
\eta_{\text{all}}=\sqrt{\alpha_{\text{p}}\cdot\overline{\eta}_{\text{p}}^{2}+\alpha_{\text{\text{t}}}\cdot\overline{\eta}_{\text{\text{t}}}^{2}+\alpha_{\text{f}}\cdot\overline{\eta}_{\text{f}}^{2}}\label{eq:overall}
\end{equation}
where $\alpha_{\text{p}},\alpha_{\text{t}},\alpha_{\text{f}}\in\left[0,1\right]$
are weighting factors with constraint $\alpha_{\text{p}}+\alpha_{\text{\text{t}}}+\alpha_{\text{f}}=1$.
The overall index $\eta_{\text{all}}$ describes the average simulation
credibility of a model from multiple assessment aspects, but it cannot
describe the worst situation. For some safety-critical systems, the
minimum index among all assessment indices is also very important
for the worst credibility requirement. The minimum credibility index
$\eta_{\text{min}}$ can be computed by
\begin{equation}
\eta_{\text{min}}=\min_{i\leq n_{\text{p}},j\leq n_{\text{t}},k\leq n_{\text{f}}}\left\{ \eta_{\text{p},i},\,\eta_{\text{t},j},\,\eta_{\text{\text{f}},k}\right\} .\label{eq:worse}
\end{equation}
In practice, a threshold $\varepsilon_{\text{min}}$ should be given
for $\eta_{\text{min}}$ to describe the actual credibility requirement.
For example, the threshold $\varepsilon_{\text{min}}=90\%$ is selected
for defining a high-credibility simulation model. If $\eta_{\text{min}}\geq\varepsilon_{\text{min}}$
is satisfied, then the overall assessment index $\eta_{\text{all}}$
can be effective for assessing the simulation accuracy.

\section{Verification and Application}

\label{sec:4}

In this section, an FPGA-based HIL simulation platform is first developed
for a multicopter system. Then, its simulation credibility is assessed
by the proposed assessment method.

\subsection{HIL Simulation Platform}

\subsubsection{Hardware Composition}

\begin{figure}
\centering \includegraphics[width=0.48\textwidth]{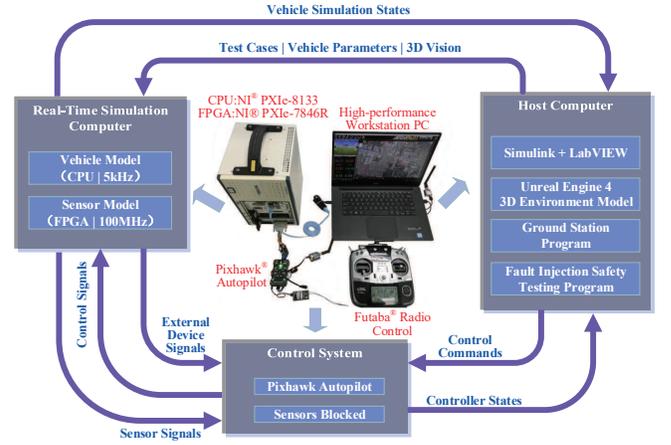}\caption{Hardware composition of the real-time HIL simulation platform.}
\label{Fig12}
\end{figure}

Based on the structure in Fig.\,\ref{Fig02}(b), an FPGA-based HIL
simulation platform is developed by the authors with the hardware
component and connection relationship shown in Fig.\,\ref{Fig12}.
The simulation computer comes from National Instruments$^{\circledR}$
(NI) with the CPU Module: PXIe-8133 (Intel Core I7 Processor, PharLap
ETS Real-Time System) and FPGA I/O Module: PXIe-7846R. The host computer
is a high-performance workstation PC with professional GPU to generate
vision data for the simulation computer. The autopilot hardware system
is the Pixhawk$^{\circledR}$ autopilot, which is a popular open-source
control system for small aircraft, vehicles, rovers, etc. All the
onboard sensors (e.g., IMU, magnetometer, and barometer) and external
sensors (e.g., GPS, accelerometer, rangefinder, and camera) of the
Pixhawk$^{\circledR}$ hardware have been blocked, and the sensor
pins are reconnected to the FPGA I/Os to generate sensor signals (interfaces:
SPI, PWM, I$^{2}$C, UART, etc.) for the control system. On the simulation
computer, the update frequency of the vehicle simulation model is
up to 5 kHz and the update frequency of sensor simulation model is
up to 100 MHz, which are fast enough for most small-scale electronic
control systems. The communication between the host computer and the
real-time simulation computer is realized by network cables with TCP
and UDP protocols.

\subsubsection{Experimental Setup}

\begin{figure}
\centering \includegraphics[width=0.48\textwidth]{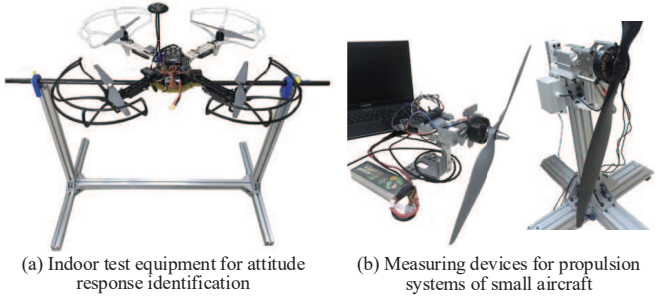}\caption{Test equipment for simulation credibility assessment.}
\label{Fig14}
\end{figure}

Based on the testing method in Fig.\,\ref{Fig08}, a series of comparative
experiments and simulations are performed to assess the simulation
credibility of the HIL simulation platform in Fig.\,\ref{Fig12}
with the proposed assessment method. The experimental setup is presented
in Fig.\,\ref{Fig14}, where an F450 quadcopter airframe (diagonal
length: 450mm, vehicle weight: 1.4kg, propulsion system: DJI E310,
battery: LiPo 3S 4000mAh) is selected as the tested system. The simulation
model of the F450 quadcopter is developed in MATLAB$^{\circledR}$/Simulink
\cite{Stevens2004AirContrl,MathWorks2019QuadcopterSim} and imported
into the HIL platform through code generation technique. In order
to test the attitude dynamics of the quadcopter, an indoor test bench
is developed with the setup shown in Fig.\,\ref{Fig14}(a), where
the quadcopter is fixed on a stiff stick (through the mass center)
with high-precision bearings to minimize friction. The quadcopter
is free to rotating along an axis smoothly, which makes it possible
to perform sweep-frequency testing for the system identification of
attitude dynamics. Fig.\,(b) presents the test benches to measure
the propulsion system parameters of the quadcopter, where the detailed
measuring methods can be found in \cite{dai2018apractical,dai2018EFF}.

\begin{figure}
\centering \includegraphics[width=0.48\textwidth]{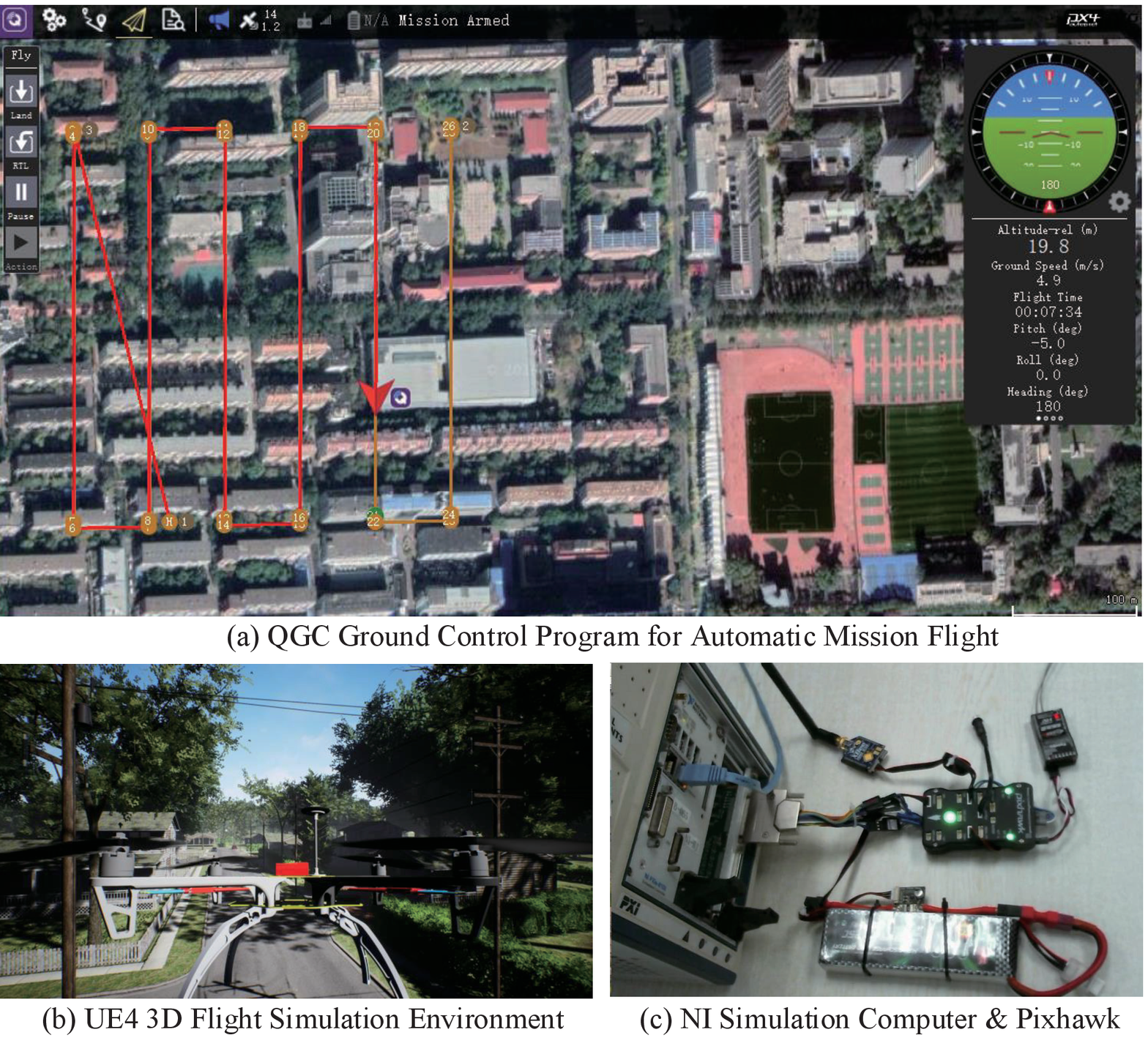}\caption{Automatic mission flight testing for F450 quadcopters. Figure (a)
presents the real-time flight trajectory observed from the ground
control station; (b) presents the high-fidelity 3D simulation scene
where a chase viewpoint is presented for observing the vehicle attitude
(the viewpoint is switchable to simulate vision from different onboard
cameras); (c) presents the real product photo of the simulation computer
and the Pixhawk$^{\circledR}$ autopilot.}
\label{Fig14-1}
\end{figure}

Since the controller parameters are associated the real aircraft and
environment, if the real control system can control the simulated
aircraft with similar flight performance as the real aircraft, the
credibility of the simulation platform can be verified indirectly.
According to our experiments, the Pixhawk$^{\circledR}$ autopilot
can control the simulated quadcopter model to finished normal flight
tasks as a real quadcopter does, where a typical automatic flight
mission test is presented in Fig.\,\ref{Fig14-1}. Besides, the HIL
simulation system is also applicable for automatic fault injection
tests which are hard to achieve in real flight tests. A video (URL:
\uline{\url{https://youtu.be/D2hIIebVXsw}}) has been released
to present the development process and testing cases for unmanned
vehicle systems with the HIL simulation platform.

\subsection{Experiments and Verification}

\subsubsection{Performance Credibility Assessment for Sensor Model}

\begin{figure}
\centering \includegraphics[width=0.45\textwidth]{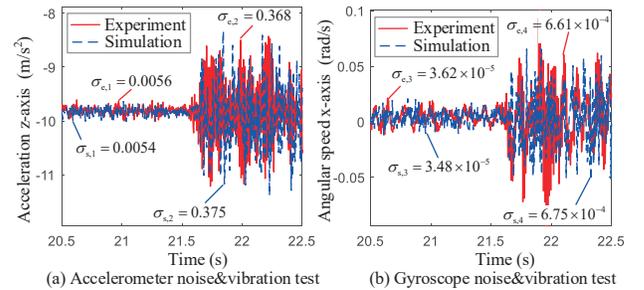}\caption{Sensor noise and vibration model verification test. The experiment
data come from a real sensor product (MPU 6000), and the simulation
data come from the simulated sensor model of the proposed real-time
HIL platform. The motor speed is stepped from 0 to 50\% to simulate
the vibration on sensor data.}
\label{Fig16-1}
\end{figure}

In order to verify the credibility of the sensor simulation methods,
experiments and simulations are performed with the testing method
presented in Fig.\,\ref{Fig08}. The obtained results are presented
in Fig.\,\ref{Fig16-1}, where the simulated accelerometer and gyroscope
are compared with the real sensor products. Figs.\,\ref{Fig16-1}(a)(b)
can verify that the sensor data generated by the HIL simulation system
are highly coincident with the sensor data on real aircraft. To further
assess the simulation effect from a quantitative view, the assessment
method proposed in Section \ref{sec:3} is carried out to assess the
simulation results presented in Fig.\,\ref{Fig16-1}.

The performance credibility index $\eta_{\text{p}}$ is selected here
because the time-domain index $\eta_{\text{t}}$ is not suitable for
analyzing stochastic signals. The standard deviation $\sigma$ is
selected as the performance parameter in (\ref{eq:Ekey}), where a
threshold $\varepsilon_{\text{p}}\approx10\%\cdot p_{\text{e}}$ is
adopted according to the measuring uncertainty as introduced in (\ref{eq:thres}).
The test results in Figs.\,\ref{Fig16-1}(a)(b) are divided into
four periods, and the simulation error for each period is obtained
by $e_{\text{p},i}=\left|\sigma_{\text{e},i}-\sigma_{\text{s},i}\right|$
according to (\ref{eq:Ekey}). Then, the simulation credibility for
each period $\eta_{\text{p},i}$ can be obtained with the results
listed in Table \ref{Tab1}. By combining the sensor credibility indices
$\eta_{\text{p},i}$ in Table \ref{Tab1}, the average simulation
credibility is obtained by (\ref{eq:aveP}) as $\overline{\eta}_{\text{p}}=94\%$.
Since the obtained credibility index $\overline{\eta}_{\text{p}}=94\%$
is far above the passing mark $\eta_{\text{pass}}=60\%$, the simulation
results can be considered accurate enough as a real sensor product.

\begin{table}
\centering{}\caption{Assessment indices obtained for sensor data.}
\label{Tab1}%
\begin{tabular}{|c|>{\centering}p{0.07\textwidth}|>{\centering}p{0.07\textwidth}|>{\centering}p{0.09\textwidth}|}
\hline 
Test period & Parameter Error $e_{\text{p},i}$ & Threshold $\varepsilon_{\text{p},i}$ & Credibility Index $\eta_{\text{p},i}$\tabularnewline
\hline 
Fig.\,\ref{Fig16-1}(a): $t$ \textless{} 21.5s & $2\times10^{-4}$ & $6\times10^{-4}$ & 91.3\%\tabularnewline
\hline 
Fig.\,\ref{Fig16-1}(a): $t$ \textgreater{} 21.5s & $7\times10^{-4}$ & $4\times10^{-3}$ & 97.4\%\tabularnewline
\hline 
Fig.\,\ref{Fig16-1}(b): $t$ \textless{} 21.5s & $1.4\times10^{-6}$ & $4\times10^{-6}$ & 90.6\%\tabularnewline
\hline 
Fig.\,\ref{Fig16-1}(b): $t$ \textgreater{} 21.5s & $1.4\times10^{-5}$ & $7\times10^{-5}$ & 96.6\%\tabularnewline
\hline 
\end{tabular}
\end{table}

\subsubsection{Frequency-domain Credibility Assessment for Attitude Dynamics}

\begin{figure}
\centering \includegraphics[width=0.45\textwidth]{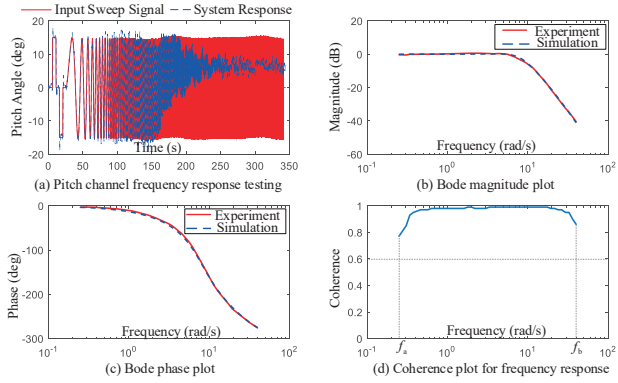}\caption{Frequency-domain assessment for the simulation fidelity.}
\label{Fig15}
\end{figure}

In order to assess the frequency-domain credibility of the simulation
platform, a series of sweeping frequency experiments are performed
by using the test bench in Fig.\,\ref{Fig14}(b). The input sweep
signal and the attitude response output are depicted in Fig.\,\ref{Fig15}(a),
where the effective frequency testing range is $\left[f_{\text{a}},f_{\text{b}}\right]=\left[0.25\text{ rad/s},40\text{ rad/s}\right]$.
The testing data in Fig.\,\ref{Fig15}(a) is processed by the software
CIFER$^{\circledR}$ \cite{remple2006aircraft}, and Figs.\,\ref{Fig15}(b)(c)(d)
present the obtained Bode's magnitude plot, Bode's phase plot, and
coherence plot, respectively. Since the coherence curve $\eta_{\text{co}}\left(f\right)$
in Fig.\,\ref{Fig15}(d) is far above the passing mark 0.6, the sweep
testing results can be considered accurate and reliable according
to the credibility criterion in (\ref{eq:Cohe}). It can be observed
from Figs.\,\ref{Fig15}(b)(c)that the errors between the experiment
curves and the simulation curves are very small in both magnitude
and phase aspects, which verify the credibility of the simulation
platform from the perspective of qualitative analysis.

In the following, the frequency-domain assessment index $\eta_{\text{f}}$
in (\ref{eq:freqIndex}) will be applied to assess the simulation
results from a quantitative perspective. For the magnitude curves
in Figs.\,\ref{Fig15}(b)(c), the magnitude credibility index is
obtained by (\ref{eq:emag})-(\ref{eq:MatFre}) as $\eta_{\text{mag}}=97.3\%$
(the average error is $e_{\text{mag}}=0.364$ and the threshold is
$\varepsilon_{\text{mag}}=2.05$); the phase credibility index is
obtained as $\eta_{\text{pha}}=97.6\%$ (the average error is $e_{\text{pha}}=2.27$
and the threshold is $\varepsilon_{\text{pha}}=13.6$). Finally, the
frequency-domain fidelity index is obtained by (\ref{eq:freqIndex})
as $\eta_{\text{f}}=97.63\%$, which indicates the pitch channel simulation
model is of high-credibility relative to the real quadcopter.

For comparison, the sweep frequency results are also analyzed by CIFER$^{\circledR}$
\cite{remple2006aircraft}, and a cost function index
$J\in[0,+\infty)$ for the modeling accuracy assessment is obtained
as $J=4.359$. Since there is no unified assessment standard, it is
hard to describe the simulation credibility only with the cost index
$J=4.359$. According to the applications, the cost index $J$ is
more suitable for comparing the simulation results obtained from the
same system, instead of comparing the simulation credibility among
different systems. Moreover, the cost index $J$ combines the magnitude
error $e_{\text{magn}}$ and phase error $e_{\text{pha}}$ by using
a constant scale factor $k_{\text{J}}$ as $J^{2}\propto(e_{\text{magn}}^{2}+k_{\text{J}}\cdot e_{\text{pha}}^{2})$,
so one of the two errors will be ignored when their orders of magnitudes
are too different (e.g., $e_{\text{magn}}\gg e_{\text{pha}}$, or
$e_{\text{magn}}\ll e_{\text{pha}}$). In summary, compared the assessment
index $J$ in CIFER$^{\circledR}$, the proposed assessment index
$\eta_{\text{f}}$ is more intuitive and efficient for the simulation
credibility assessment in the frequency domain.

\subsubsection{Time-domain Credibility for Level Flight Test}

\begin{figure}
\centering \includegraphics[width=0.42\textwidth]{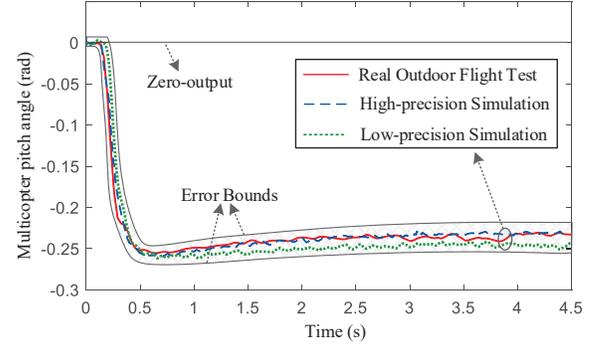}\caption{Level flight results for simulation validation. The quadcopter is
commanded to step from hovering mode to level flight mode.}
\label{Fig15-2}
\end{figure}

For quantitative analysis, more outdoor flight tests are performed
by the proposed HIL simulation platform with the typical testing results
presented in Fig.\,\ref{Fig15-2}. Since all the simulation factors
(e.g., motion, aerodynamics, sensors, and disturbances) are involved
in the level flight tests, the testing results can reflect the simulation
credibility of the proposed platform in an overall and comprehensive
way.

Three flight curves are presented in Fig.\,\ref{Fig15-2}, where
the real flight testing curve comes from the outdoor experiment, and
the high-precision simulation curve and the high-precision simulation
curve both come from the HIL simulation platform with different modeling
precision. The high-precision simulation is performed with all model
parameters being accurately measured by professional equipment or
obtained by system identification methods; the low-precision simulation
is from a simplified model with parameters obtained by analytical
estimated. It can be observed from Fig.\,\ref{Fig15-2} that the
high-precision simulation curve almost coincides with the real experimental
curve, and the low-precision simulation curve is slightly different
from the experimental curve, but the error is acceptable because it
reveals most dynamic and aerodynamic characteristics of the quadcopter.

Quantitative analysis is carried out with the results listed in Table
\ref{Tab2} to verify whether the time-domain assessment index $\eta_{\text{t}}\in(0,1]$
can distinguish the slight simulation credibility difference. For
comparison purposes, a zero-output curve (see Fig.\,\ref{Fig15-2})
is also considered as a reference for the worst simulation accuracy
case, and error bound curves (see Fig.\,\ref{Fig15-2}) are obtained
by (\ref{eq:EtThre}) as $\varepsilon_{\text{t}}\approx0.0175$. The
results in Table \ref{Tab2} demonstrate that: (i) the assessment
index is sensitive to reflect the difference among simulation platform
with different modeling precision (the high-precision model $\eta_{\text{t}}=94.4\%$
v.s. the low-precision model $\eta_{\text{t}}=73.8\%$); (ii) the
assessment index is capable of reflecting whether the accuracy satisfies
the minimum threshold (the low-precision model $\eta_{\text{t}}=73.8\%>60\%$
indicates the simulation error is acceptable); (iii) the assessment
index is sensitive to reflect the worst simulation credibility (the
zero-output curve $\eta_{\text{t}}=1.75\%\rightarrow0$).

\begin{table}
\centering{}\caption{Simulation assessment for models with different accuracy}
\label{Tab2}%
\begin{tabular}{|c|>{\centering}p{0.07\textwidth}|>{\centering}p{0.08\textwidth}|>{\centering}p{0.07\textwidth}|}
\hline 
Curve Type & Mean Error $e_{\text{t}}$ & Error Threshold $\varepsilon_{\text{t}}$ & Assessment Index $\eta_{\text{t}}$\tabularnewline
\hline 
Real Flight & 0 & 0.0175 & 100\%\tabularnewline
\hline 
High-precision & 0.0046 & 0.0175 & 94.4\%\tabularnewline
\hline 
Low-precision & 0.012 & 0.0175 & 73.8\%\tabularnewline
\hline 
Error Bounds & 0.0175 & 0.0175 & 60.0\%\tabularnewline
\hline 
Zero-output & 0.231 & 0.0175 & 1.75\%\tabularnewline
\hline 
\end{tabular}
\end{table}

\subsubsection{Overall Simulation Credibility Assessment}

With the above testing results, according to the computing expression
in (\ref{eq:aveP}), the whole performance credibility is obtained
as $\overline{\eta}_{\text{p}}=94.0\%$, the whole frequency-domain
credibility is obtained as $\overline{\eta}_{\text{f}}=97.63\%$,
and the whole time-domain credibility is obtained as $\overline{\eta}_{\text{t}}=94.4\%$.
Since the frequency-domain characteristic is usually more important
for assessing a dynamic system, by selecting weight factors $\{\alpha_{\text{p}},\alpha_{\text{t}},\alpha_{\text{f}}\}$
as $\{0.3,0.3,0.4\}$, the overall simulation credibility $\eta_{\text{all}}$
can be obtained by (\ref{eq:overall}) as $\eta_{\text{all}}=95.36\%$.
Meanwhile, the minimum simulation credibility can be obtained by (\ref{eq:worse})
as $\eta_{\text{min}}=90.6\%$. Since only several testing results
are presented in this section, the obtained indices $\eta_{\text{all}}$
and $\eta_{\text{min}}$ may be not comprehensive and representative
enough. With more testing results are considered from different angles,
the obtained indices $\eta_{\text{all}}$ and $\eta_{\text{min}}$
can become very comprehensive and representative to assess the simulation
credibility of the whole HIL system. On the other hand, these assessment
indices can help designers to find out the weak points of the simulation
models to continually improve the simulation credibility.

\section{Conclusion}

\label{sec:5}

All the above analyses demonstrate that (i) the proposed modeling
method with the FPGA-based HIL simulation system is capable of simulating
the vehicle characteristics as realistic as real vehicle systems;
(ii) the proposed simulation credibility assessment method is efficient
and practical in assessing the simulation credibility of simulation
systems. Since all proposed assessment indices are normalized to 0
to 1 and scaled to the same passing mark 0.6, we can compare and combine
different system characteristics within a unified assessment framework.
The simulation credibility assessment is important in the verification
and validation of the simulation platform compared with the real system,
which provides the basis for applying the simulation testing results
to the future safety assessment and certification frameworks, such
as the airworthiness of unmanned aircraft systems. Based on the proposed
platform and assessment method, more efficient and comprehensive automatic
testing and assessment methods will be studied in the future for electronic
systems to increase their safety and reliability levels.

\bibliographystyle{IEEEtran}
\bibliography{IEEETRO}

\end{document}